\documentclass[aps,pra,superscriptaddress,amsmath,amssymb,showpacs,twocolumn,notitlepage]{revtex4-1}
\usepackage{amsmath,amssymb,amsfonts,latexsym,color,epsfig,textcomp}
\usepackage{subfigure}
\usepackage[latin1]{inputenc}
\usepackage{dcolumn}
\usepackage{bm}
\usepackage{graphicx}
\usepackage[english]{babel}
\usepackage{booktabs}
\usepackage{multirow}
\usepackage[table,xcdraw]{xcolor}
\usepackage[colorlinks,
            linkcolor=blue,
            anchorcolor=blue,
            citecolor=blue
            ]{hyperref}

\def\bbm[#1]{\mbox{\boldmath $#1$}}

\renewenvironment{widetext@grid}{%
  \par\ignorespaces
  \setbox\widetext@top\vbox{%
   \vskip15\p@
   \hb@xt@\hsize{%
    \leaders\hrule\hfil
    \vrule\@height6\p@
   }%
   \vskip6\p@
  }%
  \setbox\widetext@bot\hb@xt@\hsize{%
    \vrule\@depth6\p@
    \leaders\hrule\hfil
  }%
  \onecolumngrid
  \let\set@footnotewidth\set@footnotewidth@ii
}{%
  \par
  \twocolumngrid\global\@ignoretrue
  \@endpetrue
}%

\begin{document}
%-----------------------------------------------------------------------------------------------------
\title{Near-field heat transfer between graphene/hBN multilayers}
%-----------------------------------------------------------------------------------------------------
\author{Bo Zhao}
\affiliation{Department of Electrical Engineering, Ginzton Laboratory, 
Stanford University, Stanford, CA 94305, USA}
\affiliation{George W. Woodruff School of Mechanical Engineering,
Georgia Institute of Technology, Atlanta, GA 30332, USA}
\author{Brahim Guizal}
\affiliation{Laboratoire Charles Coulomb (L2C), UMR 5221 CNRS-Universit\'{e} de Montpellier, F- 34095 Montpellier, France}
\author{Zhuomin M. Zhang}
\affiliation{George W. Woodruff School of Mechanical Engineering,
Georgia Institute of Technology, Atlanta, GA 30332, USA}
\author{Shanhui Fan}
\affiliation{Department of Electrical Engineering, Ginzton Laboratory, 
Stanford University, Stanford, CA 94305, USA}
\author{Mauro Antezza}
\affiliation{Laboratoire Charles Coulomb (L2C), UMR 5221 CNRS-Universit\'{e} de Montpellier, F- 34095 Montpellier, France}
\affiliation{Institut Universitaire de France - 1 rue Descartes, F-75231 Paris, France}
\date{\today}
%-----------------------------------------------------------------------------------------------------
%
%
%-----------------------------------------------------------------------------------------------------
\begin{abstract}
We study the radiative heat transfer between multilayer structures made by a periodic repetition of a graphene sheet and a hexagonal boron nitride (hBN) slab. Surface plasmons in a monolayer graphene can couple with a hyperbolic phonon polaritons in a single hBN film to form hybrid polaritons that can assist photon tunneling. For periodic multilayer graphene/hBN structures, the stacked metallic/dielectric array can give rise to a further effective hyperbolic behavior, in addition to the intrinsic natural hyperbolic behavior of hBN. The effective hyperbolicity can enable more hyperbolic polaritons that enhance the photon tunneling and hence the near-field heat transfer. However, the hybrid polaritons on the surface, i.e. surface plasmon-phonon polaritons, dominate the near-field heat transfer between multilayer structures when the topmost layer is graphene. The effective hyperbolic regions can be well predicted by the effective medium theory (EMT), thought EMT fails to capture the hybrid surface polaritons and results in a heat transfer rate much lower compared to the exact calculation.  The chemical potential of the graphene sheets can be tuned through electrical gating and results in an additional modulation of the heat transfer. We found that the near-field heat transfer between multilayer structure does not increase monotonously with the number of layer in the stack, which provides a way to control the heat transfer rate by the number of graphene layers in the multilayer structure. The results may benefit the applications of near-field energy harvesting and radiative cooling based on hybrid polaritons in two-dimensional materials.

\end{abstract}

%\pacs{\red{41.20.-q,78.67.Wj, 81.07.Oj}}
\maketitle
%-----------------------------------------------------------------------------------------------------
%41.20.-q	Applied classical electromagnetism (for submillimeter wave, microwave, and radiowave instruments and equipment, see 07.57.-c)
%78.67.Wj Graphene optical properties, 
%81.07.Oj nanoelectromechanical systems, 
%
%
%-----------------------------------------------------------------------------------------------------
\section{Introduction}\label{Intro}
%-----------------------------------------------------------------------------------------------------
As one of the fundamental modes of heat transfer, radiative heat transfer plays an important role in a wide spectrum of applications from energy harvesting to thermal management [1-5]. In the far field, the maximum radiative heat transfer rate between two objects is restricted by the black-body limit. However, if the two objects are brought very close to a distance comparable to the characteristic wavelength of the thermal radiation, the evanescent waves from each object can couple and assist photons to tunnel through the gap. This is the so-called photon tunneling and the resulted near-field heat transfer rate can be orders of magnitude larger compared to the blackbody limit [6-12]. The enhanced radiative heat transfer finds numerous applications such as thermal energy harvesting, radiative cooling, and thermal imaging [2]. Continuous efforts have been devoted to exploring new materials or structures that can result in large heat transfer rates that can benefit these applications.

Various types of surface polaritons have been extensively studied for their ability to enhance the photon tunneling and greatly boost the near-field heat transfer. Examples include surface phonon polaritons that can exist at the surface of polar dielectric materials such as SiO$_{2}$ and SiC, or surface plasmon polaritons (SPPs) that can be supported between metallic surfaces or structures [13-17]. Recently, it is demonstrated that surface plasmons in graphene can also achieve a similar role to enhance the photon tunneling between two graphene sheets [18]. Besides the surface polaritons, bulk materials constructed with periodically stacked sub-wavelength metallic and dielectric layers can also enhance near-field heat transfer. The enhancement is originated from the collective response of the multilayers that can be described based on effective medium theory. The effective dielectric function is usually anisotropic, and in some frequency range, the axial and tangential permittivities can even have opposite signs, giving rise to hyperbolic responses [19,20]. In the hyperbolic regions, the isofrequency surfaces become a hyperboloid instead of a sphere or an ellipsoid, and thus such multilayer structures can support resonance modes with unbounded tangential wavevectors. These multilayer structures have found exciting applications in sub-wavelength imaging [21-23] and near-field radiative heat transfer [24,25]. For heat transfers in particular, materials with hyperbolic responses can provide substantial enhancement of heat transfer over a broad frequency region [24].

Recently, it has been shown that the surface plasmons in graphene can couple with the phonon polaritons in hexagonal boron nitride (hBN) films to form hybrid polaritons that greatly enhance the photon tunneling [26]. The structure can yield a larger heat transfer rate than the typical polar materials such as SiO$_{2}$ and SiC. Since graphene behaves like a thin metallic layer in the structure, one could expect periodically stacked graphene and hBN film would result in a collective hyperbolic response, which may give rise to an enhanced heat transfer rate that can exceed the single-layer heterostructure. Compared to other hyperbolic metamaterials constructed with metal and isotropic dielectrics, such a multilayer structure could enable an actively tunable hyperbolic response by changing the chemical potential of graphene. Note that hBN naturally possesses two mid-infrared Reststrahlen bands that have hyperbolic response [27,28]. Thus, transitions between natural to effective hyperbolic response may also occur in such type of structures.

For these reasons, in this work, we study the near-field heat transfer between multilayers with alternating layers of graphene and hBN film. The coupled modes in hBN/graphene heterostructure are discussed and the evolution of the modes is investigated by exploring the photon tunneling probability between structures with different number of layers. The contributions of the surface polaritons are emphasized by analyzing the spectral heat flux. Predictions based on effective medium theory (EMT) is used to compare with the exact calculations to further illustrate the contributions of the surface polaritons. The heat transfer rates between structures with the same total thickness but different layers are compared.

%-----------------------------------------------------------------------------------------------------
\section{CALCULATION OF NEAR-FIELD HEAT TRANSFER}\label{HT}
%-----------------------------------------------------------------------------------------------------
%
\begin{figure}[htb]
\includegraphics[width=0.5\textwidth]{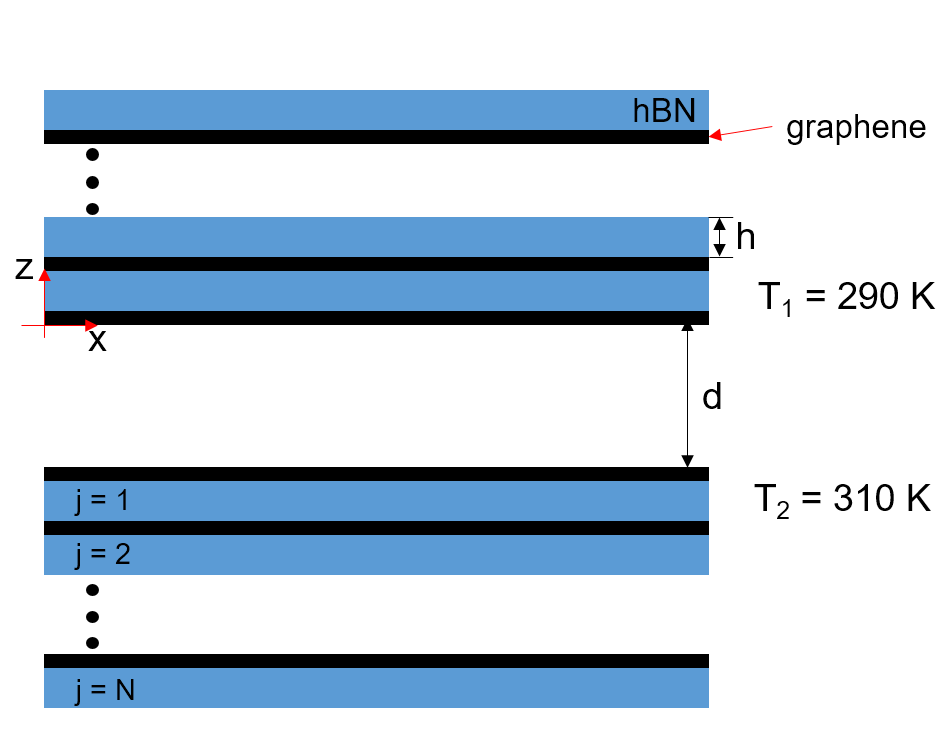}
\caption{Schematic of near-field radiative heat transfer between two graphene/hBN heterostructures.}\label{fig:schematic}
\end{figure}

Figure \ref{fig:schematic} shows the configuration of near-field radiative heat transfer between two periodic multilayer structures separated by a vacuum gap of $d$. Each period of the multilayer stack contains an hBN film with a thickness denoted as $h$ adjacent to a monolayer graphene, and the total number of layers in the structure is denoted by $N$. When $N$ = 1, the structure is simply a heterostructure containing a single layer of hBN and graphene. In Fig. \ref{fig:schematic}, the structure above the vacuum gap is the receiver with a lower temperature $T_1$ and the structure below the vacuum gap is the emitter with a higher temperature $T_2$. The temperatures are set to be $T_1$ = 290 K and $T_2$ = 310 K, respectively, and the ambient temperature is assumed to be the same as $T_1$. The heat transfer rate is calculated on the receiver [29].

\begin{figure}[tb]
\includegraphics[width=0.5\textwidth]{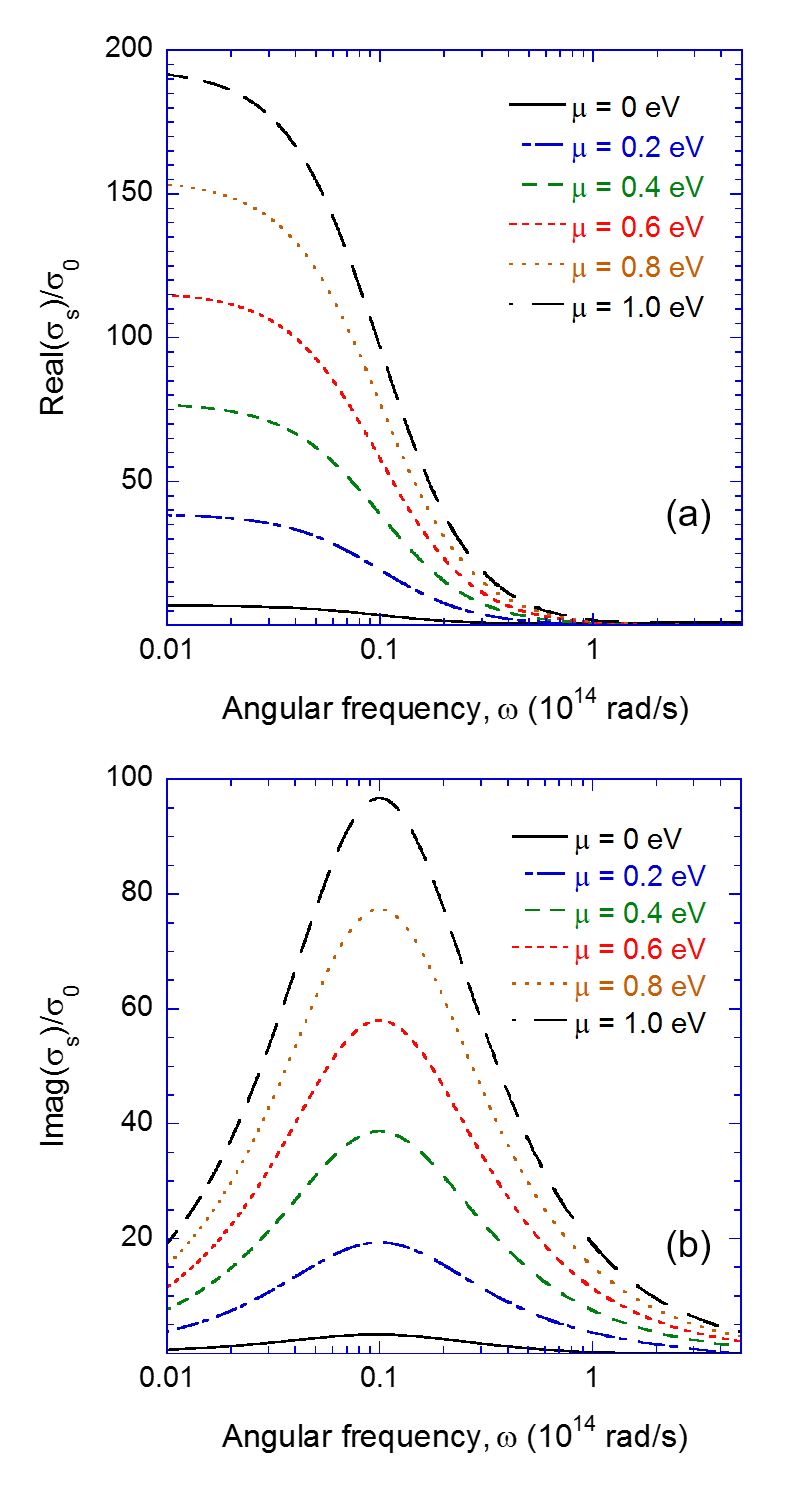}
\caption{Real and imaginary part of the sheet conductivity of graphene at different chemical potentials at $T$ = 300 K and $\tau=10^{-13}$ s. The values are normalized by $\sigma_0=e^2/(4\hbar)$. In the calculation, the properties of graphene are evaluated at the temperatures of the receiver and emitter.}\label{fig:graphene_properties}
\end{figure}

Graphene is modeled with a sheet conductivity, $\sigma_\textrm{s}$, that includes the contributions from both the intraband and interband transitions [30], i.e., $\sigma_\textrm{s}=\sigma_\textrm{D}+\sigma_\textrm{I}$, respectively [31]:

\begin{equation}\label{GrapheneConductivityD}
\sigma_\textrm{D}=\frac{i}{\omega+i/\tau}\frac{2e^2k_BT}{\pi\hbar^2}\ln\left[2\;\cosh\left(\frac{\mu}{2k_BT}\right)\right]
\end{equation}
and
\begin{equation}\label{GrapheneConductivityI}
\sigma_\textrm{I}=\frac{e^2}{4\hbar}\left[G\left(\frac{\hbar\omega}{2}\right)+i\frac{4\hbar\omega}{\pi}\int_0^{\infty}\frac{G(\xi)-G(\hbar\omega/2)}{(\hbar\omega)^2-4\xi^2}d\xi\right]
\end{equation}
where $G(\xi)=\sinh\left(\xi/k_BT\right)/\left[\cosh\left(\mu/k_BT\right)+\cosh\left(\xi/k_BT\right)\right]$. Here, $e$ is the electron charge, $\hbar$ is the reduced Planck constant, $\mu$ is the chemical potential, $\tau$ is the relaxation time and is chosen to be $10^{-13}$ s for all the calculations, $k_B$ is the Boltzmann constant, $\omega$ is the angular frequency, and $T$ is temperature that will be set to either $T_1$ or $T_2$ in the calculation depending on the location of the graphene sheets [32]. Figure \ref{fig:graphene_properties} shows the real and imaginary part of the sheet conductivity of graphene at different chemical potentials.

hBN is a uniaxial crystal in the infrared region with two mid-infrared Reststrahlen bands. We assume its optical axis is in the $z$-direction for the structure in Fig. \ref{fig:schematic}. The in-plane and out-of-plane dielectric functions include the contribution from the in-plane phonon vibrations ($\omega_{\textrm{TO}, \bot}$ = 1370 $\textrm{cm}^{-1}$ and $\omega_{\textrm{LO},\bot}$ = 1610 $\textrm{cm}^{-1}$) and out-of-plane phonon vibrations ($\omega_{\textrm{TO},\parallel}$ = 780 $\textrm{cm}^{-1}$ and $\omega_{\textrm{LO},\parallel}$ = 830 $\textrm{cm}^{-1}$), respectively, as given by
\begin{equation}\label{hBN_properties}
\epsilon_m=\epsilon_{\infty,m}\left(1+\frac{\omega^2_{\textrm{LO},m}-\omega^{2}_{\textrm{TO},m}}{\omega^{2}_{\textrm{TO},m}-i\gamma_m\omega-\omega^2}\right)
\end{equation}
where $m=\,\parallel, \bot$[33] denote either the out-of-plane or the in-plane directions, respectively. The other parameters used are $\epsilon_{\infty,\parallel}=2.95$, $\gamma_{\parallel}=4\,\textrm{cm}^{-1}$, $\epsilon_{\infty,\bot}=4.87$, $\gamma_{\bot}= 5\,\textrm{cm}^{-1}$. Due to the small damping coefficients as compared to the phonon frequencies, the in-plane and out-of-plane dielectric functions of hBN possess opposite signs in the Reststrahlen bands, making hBN a natural hyperbolic material. 

The near-field radiative heat flux $q$ is calculated based on fluctuational electrodynamics [3]
\begin{equation}\label{NFHT}
q=\frac{1}{4\pi^2}\int_0^\infty\left[\Theta\left(\omega,T_2\right)-\Theta\left(\omega,T_1\right)\right]\left[\int_0^\infty\xi\left(\omega,\beta\right)\beta d\beta\right]d\omega
\end{equation}
where $\Theta\left(\omega,T\right)$ is the average energy of a Planck oscillator, $\beta$ designates the magnitude of the wave vector in the $x-y$ plane, and $\xi\left(\omega,\beta\right)$ is the photon tunneling probability (also called energy transmission coefficient). If the integration is done over $\beta$ only, the result is the spectral heat flux. The photon tunneling probability includes contributions of both the transverse electric (TE) waves (or $s$-polarization) and transverse magnetic (TM) waves (or $p$-polarization), that is, $\xi\left(\omega,\beta\right)=\xi_s\left(\omega,\beta\right)+\xi_p\left(\omega,\beta\right)$. Each polarization contains the contribution from the propagating ($\beta<k_0$) and evanescent waves ($\beta>k_0$), where $k_0=\omega/c_0$ is the magnitude of the wave vector in vacuum and $c_0$ is the speed of light in vacuum [29]:
\begin{widetext}
\begin{equation}\label{transcoeff}
\xi_j\left(\omega,\beta\right)=\left\{
\begin{array}{cr}
\frac{\left(1-|r_{1j}|^2\right)\left(1-|r_{2j}|^2\right)
-|t_{1j}|^2\left(1-|r_{2j}|^2\right)-|t_{2j}|^2\left(1-|r_{1j}|^2-|t_{1j}|^2\right)}
{|1-r_{1j}r_{2j}e^{2ik_{z0}d}|^2}, $$ \;\beta<k_0\\
\frac{4\left[\textrm{Im}\left(r_{1j}\right)\textrm{Im}\left(r_{2j}\right)\right]e^{-2|k_{z0}|d}}{|1-r_{1j}r_{2j}e^{2ik_{z0}d}|^2}, $$ \;\beta>k_0
\end{array}
\right.
\end{equation}
\end{widetext}
where $j$ is for either $s$ or $p$ polarization, 1 and 2 respectively denote the receiver and emitter, $r$ and $t$ are respectively the corresponding reflection and transmission coefficients, $k_{z0}$ is the $z$-component of the wave vector in vacuum, and Im takes the imaginary part [34]. 

For structure with $N = 1$, the reflection and transmission coefficients for both $s$- and $p$-polarizations take the following forms [3,26]
\begin{equation}\label{r}
r=r_{12}+\frac{t_{12}t_{21}r_{23}e^{i2k_{z,2}h}}{1-r_{21}r_{23}e^{i2k_{z,2}h}}
\end{equation}
and 

\begin{equation}\label{t}
t=\frac{t_{12}t_{23}e^{ik_{z,2}h}}{1-r_{21}r_{23}e^{i2k_{z,2}h}}
\end{equation}

where 1, 2, and 3 are the indexes for the vacuum region above hBN film, the hBN film region, and the vacuum region below hBN film, respectively. Also, 

\begin{equation}\label{rabs}
r_{ab,s}=\frac{k_{z,a}-\sigma_\textrm{s}\omega\mu_0-k_{z,b}}{k_{z,a}+\sigma_\textrm{s}\omega\mu_0+k_{z,b}}
\end{equation}
\begin{equation}\label{tabs}
t_{ab,s}=\frac{2k_{z,a}}{k_{z,a}+\sigma_\textrm{s}\omega\mu_0+k_{z,b}}
\end{equation}
\begin{equation}\label{rabp}
r_{ab,p}=\frac{k_{z,a}\epsilon_{\bot,b}-k_{z,b}\epsilon_{\bot,a}+k_{z,a}k_{z,b}\frac{\sigma_\textrm{s}}{\omega\epsilon_0}}{k_{z,a}\epsilon_{\bot,b}+k_{z,b}\epsilon_{\bot,a}+k_{z,a}k_{z,b}\frac{\sigma_\textrm{s}}{\omega\epsilon_0}}
\end{equation}
\begin{equation}\label{tabp}
t_{ab,p}=\frac{2k_{z,a}\epsilon_{\bot,b}}{k_{z,a}\epsilon_{\bot,b}+k_{z,b}\epsilon_{\bot,a}+k_{z,a}k_{z,b}\frac{\sigma_\textrm{s}}{\omega\epsilon_0}}
\end{equation}
Here, $a$ and $b$ can be 1, 2, or 3, and $\epsilon_0$ is the vacuum permittivity, $\mu_0$ is the vacuum permeability. Note that the $z$-component of the wavevector in a given region takes different form depends on the polarization. For $s$-polarization, $k_{z,a}=\left(\epsilon_{\bot,a}k^2_0-\beta^2\right)^{1/2}$, and for $p$-polarization, $k_{z,a}=\left(\epsilon_{\bot,a}k^2_0-\epsilon_{\bot,a}\beta^2/\epsilon_{\parallel,a}\right)^{1/2}$. For regions with an isotropic medium like regions 1 and 3, $\epsilon_1=\epsilon_3=\epsilon_{\bot}=\epsilon_{\parallel}=1$. On the interface without graphene, $\sigma_\textrm{s}=0$. For $N>1$, the reflection and transmission coefficients can be obtained by modifying the reflection and transmission coefficient at the interface between the hBN film and the lower vacuum in Eqs. (\ref{r}) and (\ref{t}). For example, the structure with $N=2$ has an additional graphene layer and an hBN film added below and above the structure with $N=1$. Therefore, $r_{23}$ is not the reflection coefficient between the interface of two media described by Eqs. (8) and (10), but takes a form that is the same with Eq. (6), excepting that region 1 becomes hBN. The transmission coefficient, $t_{23}$, can be modified in a similar way and takes a form that is the same with Eq. (7) with region 1 being hBN. This process can be repeated to obtain the reflection and transmission coefficients for structures with $N$ layers, and the results are cross-checked using a scattering matrix method [35]. There is an alternative method in which graphene is modeled as a layer of thickness $\Delta$ = 0.3 nm with an effective dielectric function $\epsilon_\textrm{eff,G}=1+i\sigma_\textrm{s}/\left(\epsilon_0\omega\Delta\right)$ [36]. Both methods yield essentially identical results with less than 0.5$\%$ in the predicted total heat flux [26]. The calculations in this work are all based on the above-mentioned analytical expressions. The latter treatment can facilitate the understanding of the physical mechanism, as will be discussed in the following.

%-----------------------------------------------------------------------------------------------------
\section{HYBRID POLARITONS IN GRAPHENE/HBN MULTILAYERS }
%-----------------------------------------------------------------------------------------------------
%
\begin{figure}[htb]
\includegraphics[width=0.5\textwidth]{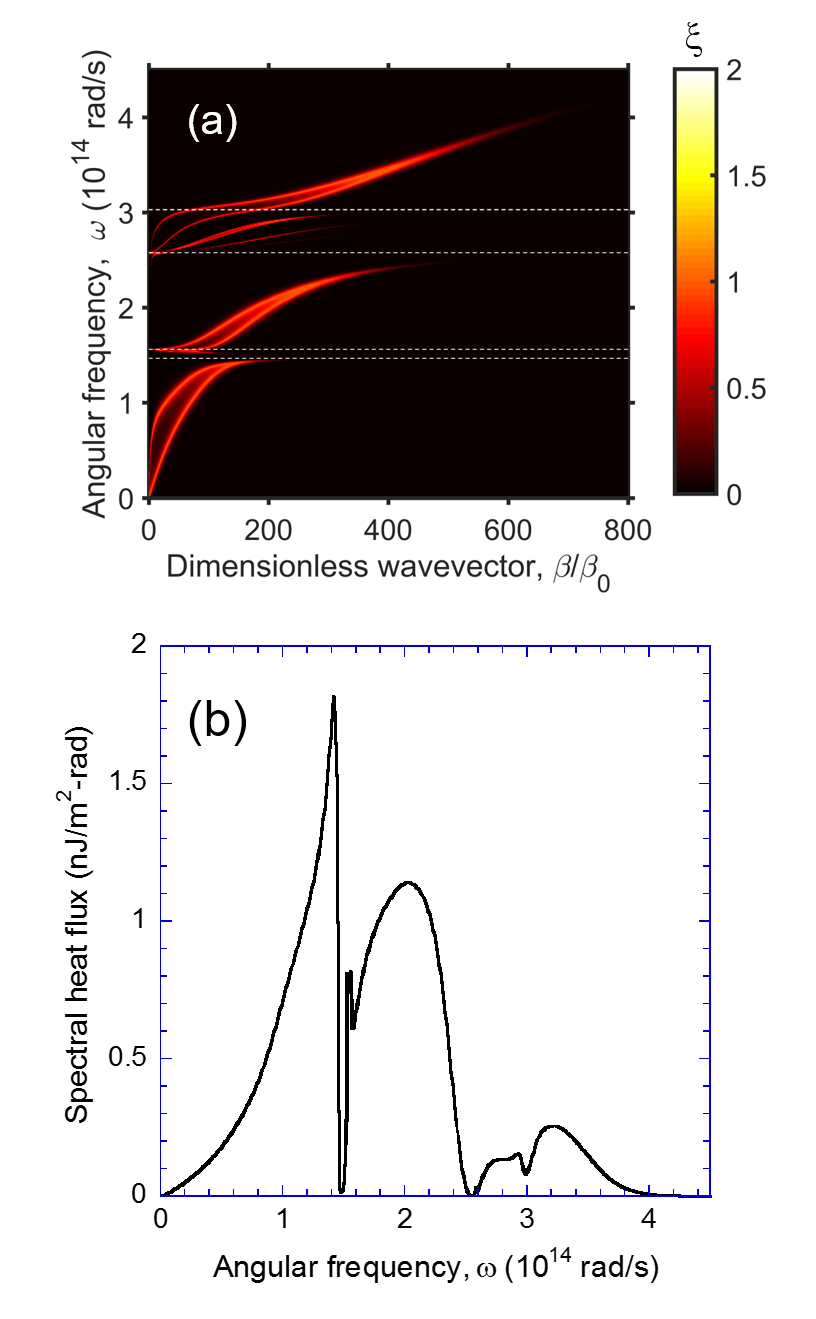}
\caption{(a) Photon tunneling probability contour and (b) spectral heat flux between two graphene/hBN heterostructures with $N$ = 1. The dashed lines indicate the two Reststrahlen bands of hBN. The parameters are $d$ = 20 nm, $h$ = 50 nm, and $\mu$ = 0.3 eV.}\label{fig:N_1}
\end{figure}

Figure \ref{fig:N_1}(a) demonstrates the photon tunneling probability contours for graphene/hBN heterostructure that contains one unit cell on either side of the vacuum gap ($N$ = 1). The wavevector is normalized using $\beta_0=\omega_0/c_0$ with $\omega_0=1\times10^{14}$ rad/s. Similar to the observations in Ref. [26], when the structure contains only one layer of hBN covered by graphene, hybrid polaritons are formed due to the coupling between hyperbolic phonon polaritons (HPPs) and surface plasmons in graphene. The bright bands indicate the excitation of the hybrid polaritons that enables a high probability of photon tunneling. The polaritons inside the two Reststrahlen bands of hBN are hyperbolic plasmon-phonon polaritons (HPPPs). HPPPs preserve the hyperbolic-waveguide-mode features as in an uncovered hBN film, and they have opposite group velocities in the two Reststrahlen bands. The hybrid polaritons outside the hyperbolic regions are surface plasmon-phonon polaritons (SPPPs), which are surface modes featured with a strong localized field on the interface with graphene. It can be seen from Fig. 3(b) that the SPPPs are the major contribution of the radiative heat transfer, which is 178 kW/m$^2$. There are two branches of SPPPs outside the hyperbolic regions, while the number of branches of HPPPs is affected by the thickness of hBN film. For an infinite thick substrate, the HPPPs merge to form a continuous band.

\begin{widetext}

\begin{figure}[hbt]
\includegraphics[width=0.75\textwidth]{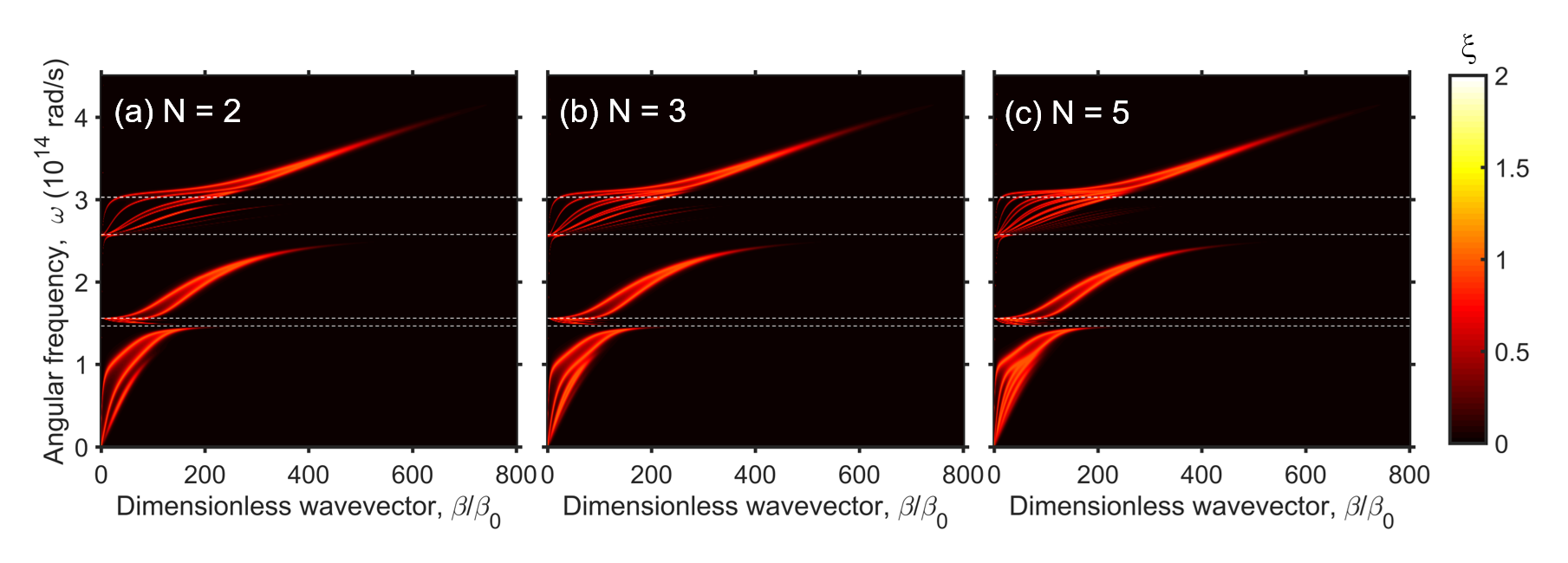}
\caption{(a) Photon tunneling probability contour and (b) spectral heat flux between two graphene/hBN heterostructures with $N$ = 1. The dashed lines indicate the two Reststrahlen bands of hBN. The parameters are $d$ = 20 nm, $h$ = 50 nm, and $\mu$ = 0.3 eV.}\label{fig:N_2_3_5}
\end{figure}

\end{widetext}

\begin{figure}[hbt]
\includegraphics[width=0.5\textwidth]{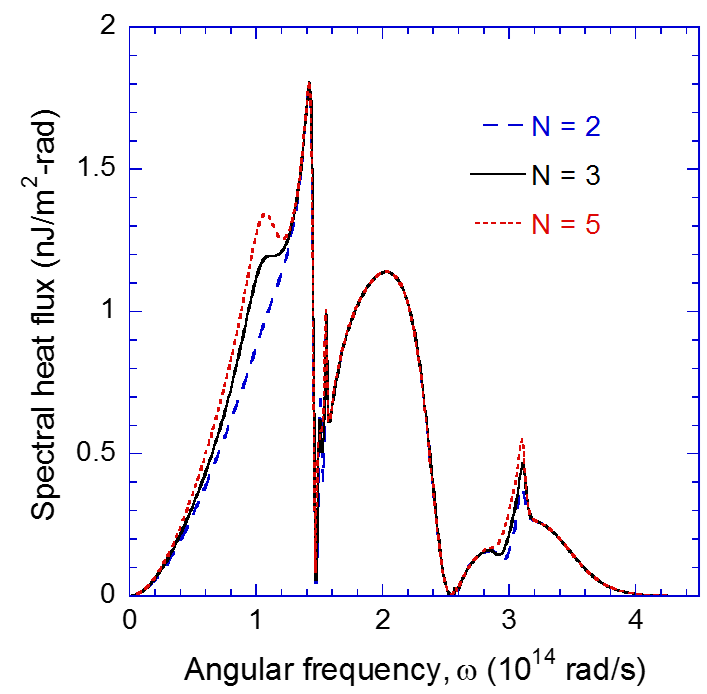}
\caption{Spectral heat flux between two graphene/hBN heterostructures with  $N$  = 2, 3, and 5. The parameters are $d$ = 20 nm, $h$ = 50 nm, and $\mu$ = 0.3 eV.}\label{fig:spectrum235}
\end{figure}

As the number of layers in the structure increases, more hybrid modes occur as indicated in Figs. \ref{fig:N_2_3_5}(a), (b), (c), in which $N$ = 2, 3, and 5, respectively. The heat transfer rate also increases to 196 kW/m$^2$, 208 kW/m$^2$, and 218 kW/m$^2$, respectively. The origin of the increase can be seen clearly from the spectral heat flux for the three cases in Fig. 5. As indicated in Fig. 4, the number of polaritons branches below the lower Reststrahlen band increases as $N$ increases. These additional bands result in a higher spectral heat flux around 10$^{14}$ rad/s as shown in Fig. \ref{fig:spectrum235}. The HPPPs extend to the frequencies higher than the upper hyperbolic region of hBN, making the number of the total branches equal to 2$N$. These additional branches do not extend to large frequencies like SPPPs but are bounded within a certain frequency region, and they also lead to a higher spectral heat flux around 3.1 $\times\,10^{14}$ rad/s. The spectral heat flux does not have a noticeable change other than the two mentioned spectral ranges. The observations can be understood by considering the effective behavior of the multilayers based on effective medium theory (EMT), which suggests a uniform property described as [37]

\begin{eqnarray}\label{EMTper}
\epsilon_{\bot,\textrm{EMT}}=f\epsilon_{\bot,\textrm{G}}+(1-f)\epsilon_{\bot,\textrm{hBN}}\nonumber \\
\epsilon_{\parallel,\textrm{EMT}}=\left(\frac{f}{\epsilon_{\parallel,\textrm{G}}}+\frac{1-f}{\epsilon_{\parallel,\textrm{hBN}}}\right)^{-1}
\end{eqnarray}
in which
\begin{equation}\label{f}
f=\frac{\Delta}{\Delta+h}
\end{equation}%
\begin{figure}[hbt]
\includegraphics[width=0.5\textwidth]{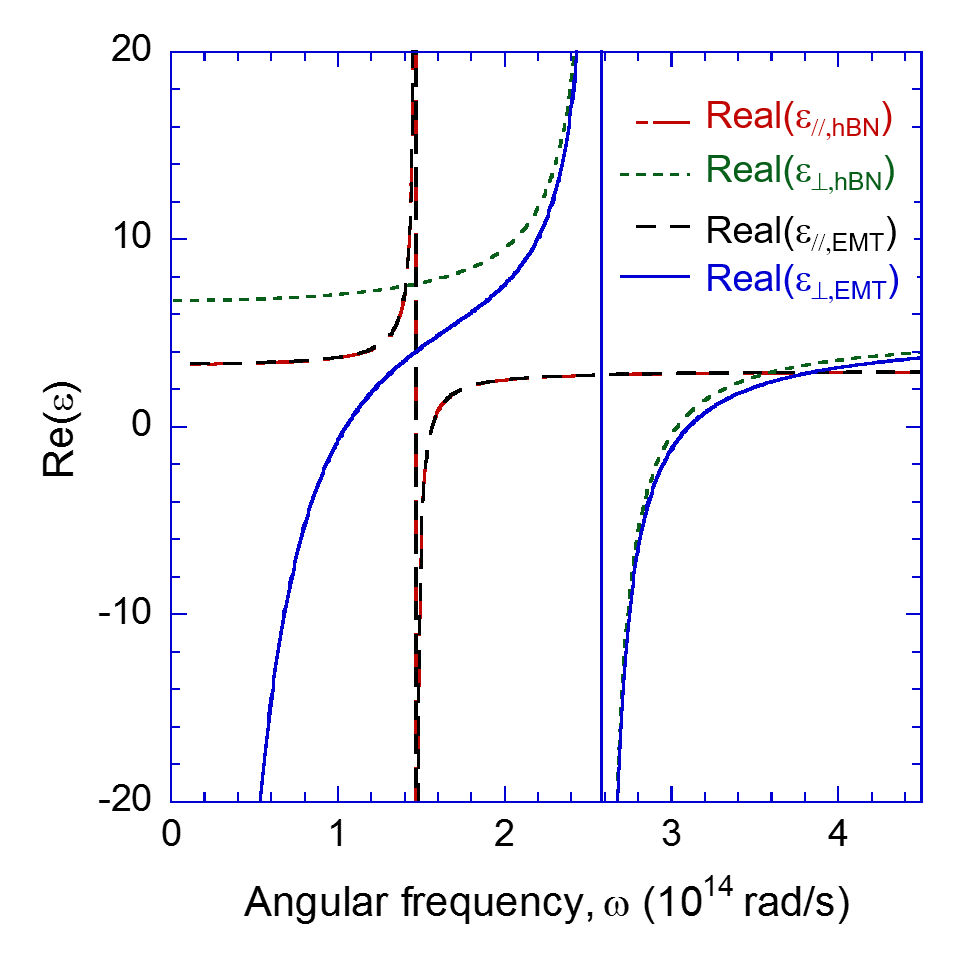}
\caption{Dielectric functions of hBN and the effective dielectric functions of the multilayer structure based on Eq. (12). The parameters are $h$ = 50 nm and $\mu$ = 0.3 eV.}\label{fig:EMT}
\end{figure}

\begin{figure}[hbt]
\includegraphics[width=0.5\textwidth]{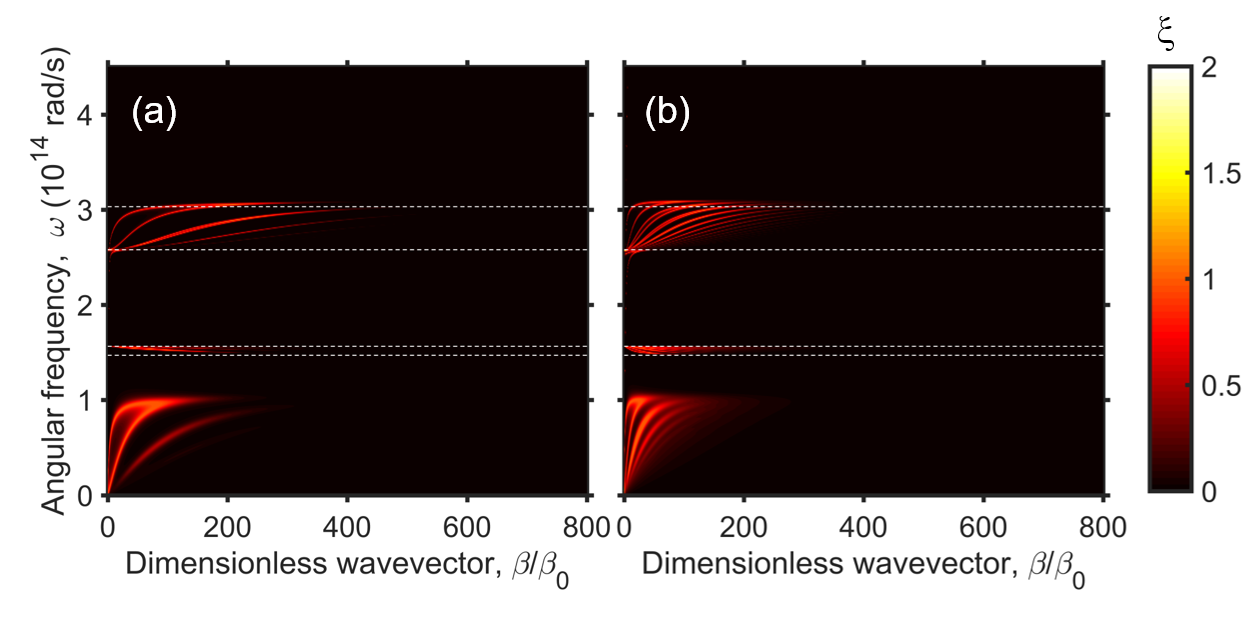}
\caption{Photon tunneling probability contours between two graphene/hBN multilayer structures calculated based on the effective properties. The total thickness of the structure is (a) 50 nm and (b) 250 nm, corresponding to the cases shown in Fig. 3(a) and Fig. 4(c), respectively. }\label{fig:EMTcontour}
\end{figure}
is the filling fraction. Note that the optical axis of the multilayer is still in the $z$-direction. Figure \ref{fig:EMT} shows the dielectric functions of hBN and the effective dielectric functions of the multilayer structure based on Eq. (12) using $\epsilon_{\bot,\textrm{G}}=\epsilon_{\parallel,\textrm{G}}=\epsilon_{\textrm{eff,G}}$. Compared to hBN, EMT predicts a very similar out-of-plane dielectric function, but the in-plane dielectric function is modified significantly due to the metallic behavior of graphene, especially at long wavelengths. Thus, there is a new hyperbolic region formed below the lower Reststrahlen band from 0 to 1.05 $\times$ 10$^{14}$ rad/s. The high-frequency bond of the upper hyperbolic region of hBN is extended to a slightly higher frequency from 3.03 $\times$ 10$^{14}$ rad/s to 3.1 $\times$ 10$^{14}$ rad/s. These changes correspond well with the observations in Fig. 4. The multiple bands in the lower frequency region are bonded by 1.05 $\times$ 10$^{14}$ rad/s and exhibit a dispersion similar to HPPPs. In the frequency region between 1.05 $\times$ 10$^{14}$ rad/s to $\omega_{\textrm{TO},\parallel}$, SPPPs are still present due to the lack of hyperbolicity. Meanwhile, due to the extension of the hyperbolic region, HPPPs in the upper Reststrahlen band extend to a slightly higher frequency to 3.1 $\times$ 10$^{14}$ rad/s. Therefore, EMT provides a qualitative explanation to understand the polariton bands in the multilayer structures. Surprisingly, the effect of the multilayer that creates effective hyperbolic regions can be observed even when $N$ = 2. The thickness of the hBN film can be changed so that the hyperbolic regions can be tuned.

\begin{figure}[hbt]
\includegraphics[width=0.5\textwidth]{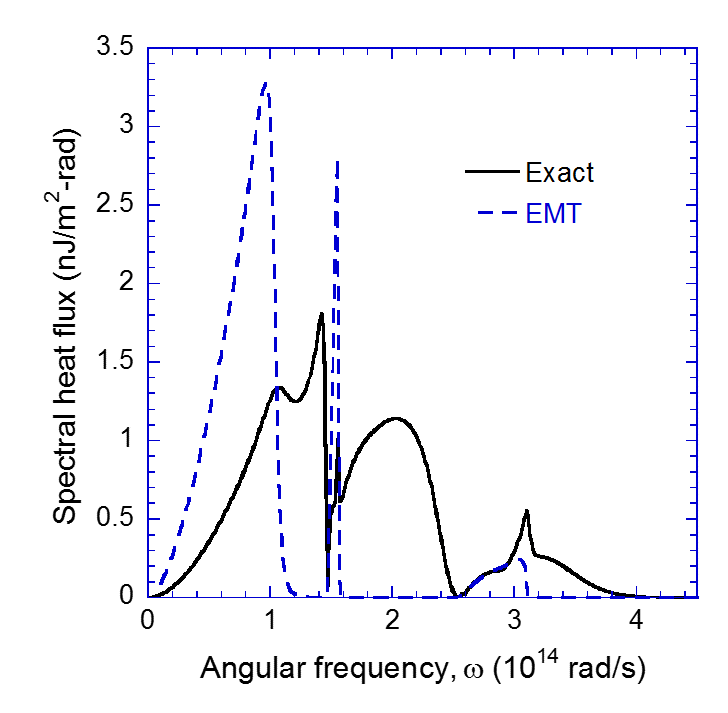}
\caption{Spectral heat flux between a multilayer structure and its mirror image calculated based on exact formula and EMT. The geometry of the structure is the same as the case in Fig. 4(c).}\label{fig:ExactEMT}
\end{figure}

Although EMT gives a qualitative explanation, it cannot capture the details of the polaritons. Figures \ref{fig:EMTcontour}(a) and (b) describe such an effect where the effective properties are used and the thickness of the structure is 50 nm and 250 nm, respectively, corresponding to the cases shown in Fig. 3(a) and Fig. 4(c). The hyperbolic polaritons are well captured by the EMT, though the dispersions are different. The SPPPs, however, do not show up. This can be understood since the mode profile of SPPPs is largely confined on the surface with graphene [33], and this inherent inhomogeneity is not captured by EMT. The effect can be better seen from Fig. \ref{fig:ExactEMT}, which displays the spectral heat flux calculated based on EMT and exact formula for the $N$ = 5 case. The EMT does not capture the peaks in the frequency region where SPPPs exist. Since the heat flux is mainly contributed by the SPPPs, EMT yields a much lower heat flux $q$ = 176 kW/m$^2$ compared to the $q$ = 218 kW/m$^2$. This indicates the importance of the first layer in enhancing the photon tunneling and near-field heat transfer. In fact, exact calculation shows that if the graphene on the top is removed, the SPPPs bands disappear and $q$ drops to 30 kW/m$^2$. The contour plot for $\xi$ looks similar to Fig. 7, though not shown here. Note that surface polaritons also play a critical role in the near-field heat transfer between hyperbolic metamaterials [25,38]. Thus, it can be concluded that EMT is not valid when the surface polaritons dominate the heat transfer.

\begin{widetext}

\begin{figure}[hbt]
\includegraphics[width=0.75\textwidth]{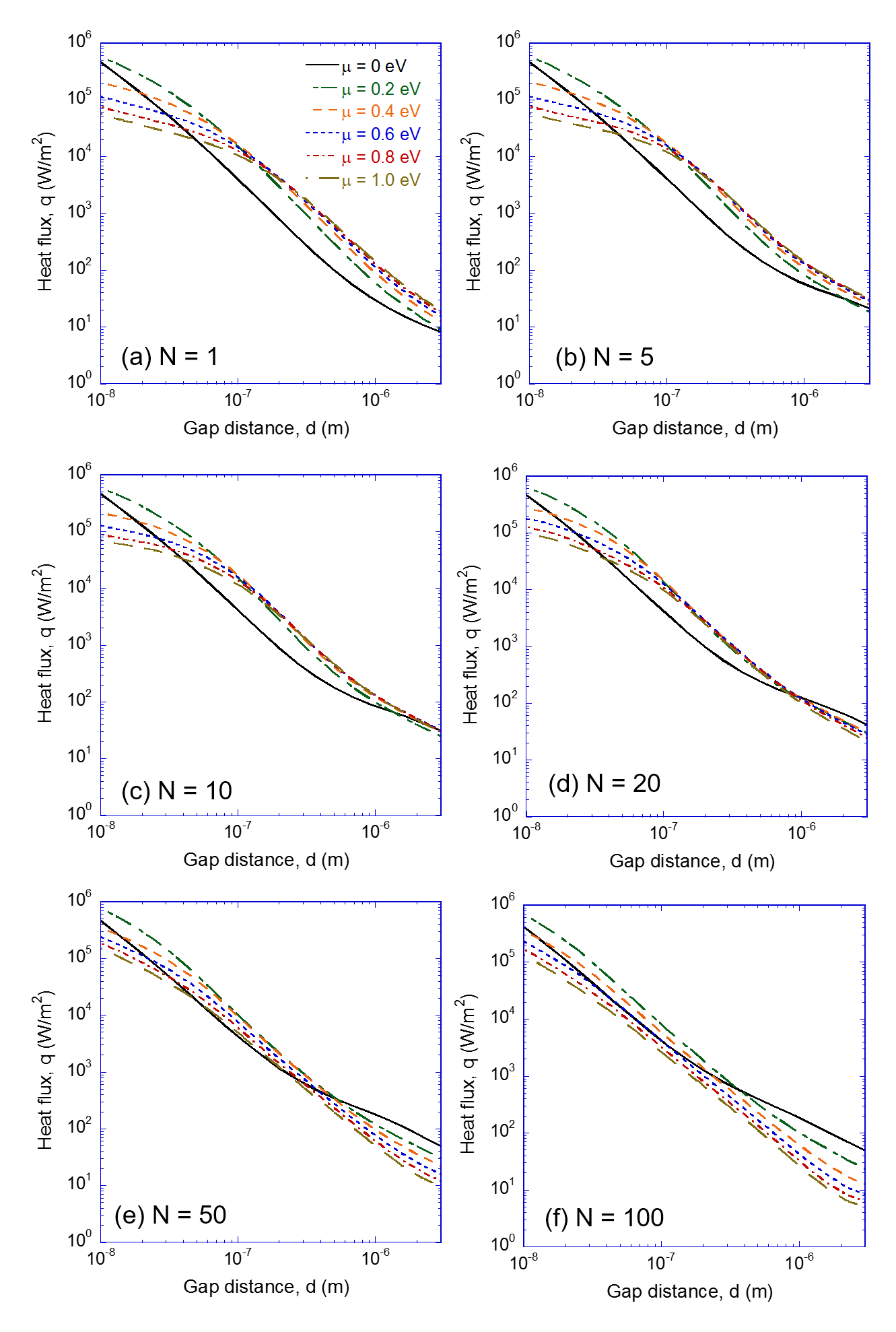}
\caption{Heat flux between structures with different layers at different graphene chemical potentials. The total thickness of the structure is fixed at 1 $\mu$m. }\label{fig:ALL}
\end{figure}

\end{widetext}

As shown in the example above, more than half of the contribution to the heat transfer arises from the graphene layers that are immediately adjacent to the vacuum gap, with additional contributions arise from the hyperbolic behavior of the multilayer structures. This observation is in general in agreement with a recent investigation of  heat transfer between hyperbolic metamaterials, which showed that a single-layer structure could operate as well or better than hyperbolic metamaterials especially in the limit of small vacuum gap sizes [39]. With the gap sizes that we consider here, there can be significant contributions from both the multi-layers and the top surfaces.

%-----------------------------------------------------------------------------------------------------
\section{EFFECT OF CHEMICAL POTENTIAL AND NUMBER OF LAYERS }\label{Parameter}
%-----------------------------------------------------------------------------------------------------

In this section, we compare the heat transfer rate for single-layer structure and multilayer structure at different chemical potentials, in order to offer a guidance to choose the optimized chemical potential and number of layers to achieve maximum heat flux. Figure \ref{fig:ALL} demonstrates the heat transfer rate between two identical structures at different chemical potentials chosen from a set of 0, 0.2, 0.4, 0.6, 0.8 and 1.0 eV, and $N$ is taken from a set of 1, 5, 10, 20, 50, and 100. The total thickness of the structures is fixed at 1 $\mu$m and thus $h$ is one micrometer divided by $N$. At small gap distances, the near-field heat transfer rates are much larger than the black-body limit, i.e., 123 W/m$^2$. Compared to other chemical potentials, $\mu$ = 0.2 eV yields the largest heat transfer rate for all $N$ when $d$ is smaller than 100 nm. The maximum heat transfer rate at $d$ = 20 nm is 303 kW/m$^2$ when $N$ = 50. This rate is much larger compared to that between polar materials (such as SiC and SiO2, yielding 46 kW/m$^2$ and 138 kW/m$^2$, respectively) and plasmonic materials (like heavily doped-Si, yielding 44 kW/m$^2$ at a doping level of 10${^{19}}$ cm$^{-3}$ [15]) that are known to yield very large heat transfer rates.

\begin{figure}[hbt]
\includegraphics[width=0.5\textwidth]{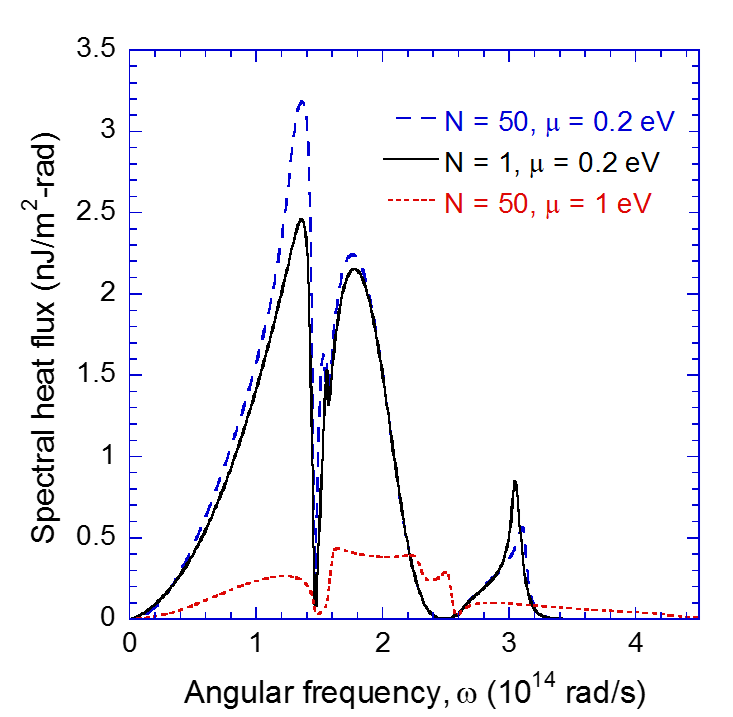}
\caption{Heat flux between structures with different layers and graphene chemical potential. The total thickness of the structure is fixed at 1 $mu$m and $h$ is one micrometer divided by $N$.}\label{fig:Typicalexample}
\end{figure}

\begin{widetext}

\begin{figure}[hbt]
\includegraphics[width=0.75\textwidth]{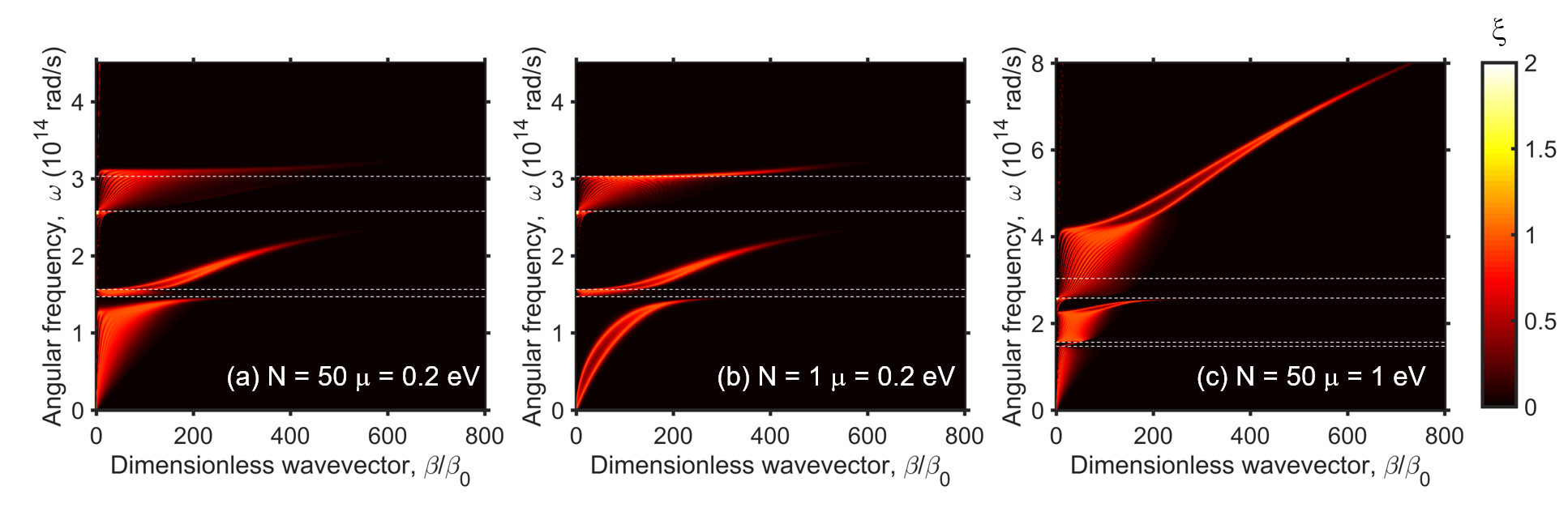}
\caption{Photon tunneling probability contours between structures with different layers and chemical potential. The total thickness of the structure is fixed at 1 $\mu$m and $h$ is one micrometer divided by $N$.}\label{fig:Typicalexamplecontour}
\end{figure}

\end{widetext}

The spectral heat flux corresponding to the maximum heat rate at $d$ = 20 nm (i.e., $N$ = 50 and $\mu$ = 0.2 eV) is shown in Fig. \ref{fig:Typicalexample}. The cases when $N$ = 1 or $\mu$ = 1 eV are also shown in comparison, in which case $q$ = 270 kW/m$^2$ and 68 kW/m$^2$, respectively. The corresponding photon tunneling probability plots are displayed in Fig. 11. The spectral heat flux spectra for the cases with $\mu$ = 0.2 eV are very similar. The majority of the heat flux is contributed by the polaritons below the lower Reststrahlen band and the SPPPs between the two Reststrahlen bands. Increasing the number of graphene layers allows more HPPPs inside the effective hyperbolic regions below the lower Reststrahlen band and above the higher Reststrahlen band as shown in Fig. \ref{fig:Typicalexamplecontour}(a), resulting in a higher spectral heat flux in the corresponding frequency regions. The wavevectors of the multiple bands of HPPPs in the hyperbolic regions are very closely spaced and they eventually form a continuous region if $N$ goes to infinity. Note that a larger $N$ does not always result in a larger $q$. For $\mu$ = 0.2 eV, $q$ would decrease to 267 kW/m$^2$ if $N$ = 100. This value is even smaller than the single layer structure with the same total thickness. Thus, one can expect there is a layer number that can maximize the heat transfer, and the number of graphene layer can be used to design a structure that yields a certain heat transfer rate. 

The heat transfer rate at $\mu$ = 0.2 eV is relatively large since smaller $\mu$ allows the SPPPs to extend to larger wavevectors [26] as can be seen from Fig. 11(a). It can be seen from Fig. 9 that the heat transfer rate at $\mu$ = 0.2 eV yields the largest $q$ for structures with different $N$. The largest heat transfer rate is achieved around 0.1 eV rather than 0 eV because interband transitions dominate $\sigma_\textrm{s}$ in the near-infrared region at $\mu$ = 0 eV and graphene does not support surface plasmons in the wavelength range of interest [40]. When $\mu$ = 1 eV, graphene plasmons expand to cover a wider frequency range as indicated in Fig. 11(c). However, due to the frequency dependence of $\Theta$, the high-frequency SPPPs do not contribute to the spectral heat flux significantly. The polaritons in the lower frequency region do not extend to large wavevectors due to the high $\mu$, and thus the spectral heat flux decrease drastically. The high chemical potential also makes the lower Reststrahlen band become an effective metallic region without a hyperbolicity, which can be seen from the disappearance of the multiple HPPPs bands. For higher emitter temperatures, increasing the chemical potential may result in a larger heat transfer rate since the high-frequency polaritons would be more significant. The strong dependence of the near-field heat transfer on the chemical potential offers another way to actively tune near-field heat transfer besides changing $N$ [41,42]. Note that after the submission of this paper, a paper studying similar system appeared [43].
%

%-----------------------------------------------------------------------------------------------------
\section{CONCLUSIONS }\label{CONCLUSIONS}
%-----------------------------------------------------------------------------------------------------
In conclusion, the multilayer structures consisting of graphene and hBN film enable more hybrid polaritons compared to the single-layer structure. HPPPs can be supported in the effectively formed hyperbolic regions that are different from the original hyperbolic regions of hBN. The majority of the near-field heat transfer, however, is still contributed by SPPPs when graphene is the topmost layer. EMT can predict the effective hyperbolic regions but fails to capture the surface polaritons, and thus yields a much lower heat transfer rate compared to the exact calculations. In additional to actively changing the graphene chemical potential, the near-field heat transfer can also be modulated through the number of graphene layers in the structure, which changes the number of the polariton bands. The results demonstrate the possibility to construct hyperbolic metamaterials with two-dimensional materials and may benefit the applications such as near-field energy harvesting and thermal imaging.

\textbf{Acknowledgement}
B. Z. and S. F. acknowledges the support from Advanced Research Projects Agency-Energy (ARPA-E), IDEAS program (project title: Demonstration of Near-Field Thermophotovoltaic (TPV) Energy Generation), as well as the support from the Global Climate and Energy Project at Stanford University, and the Department of Energy ``Material Interactions in Energy Conversion$\textquotedblright$ Energy Frontier Research Center under Grant No. DE-SC0001293. Z.M.Z. acknowledges the support from the National Science Foundation (CBET-1603761).

%

%

%-----------------------------------------------------------------------------------------------------
% BIBLIO
%-----------------------------------------------------------------------------------------------------

\end{document}